# Mapping the Diffuse Ultraviolet Sky with GALEX

**Jayant Murthy[1*], R. C. Henry[2] & N. V. Sujatha[1]**

[1]Indian Institute of Astrophysics, Bengalooru 560 034, India, [2]Dept. of Physics and Astronomy, The Johns Hopkins University, Baltimore, MD. 21218, USA

*Correspondence address: jmurthy@yahoo.com

## *Abstract*

We present a map of the diffuse ultraviolet cosmic background in two wavelength bands (FUV: 1530 Å; NUV: 2310 Å) over almost 75% of the sky using archival data from the GALEX mission. Most of the diffuse flux is due to dust-scattered starlight and follows a cosecant law with slopes of 545 photons cm$^{-2}$ s$^{-1}$ sr$^{-1}$ Å$^{-1}$ and 433 photons cm$^{-2}$ s$^{-1}$ sr$^{-1}$ Å$^{-1}$ in the FUV and NUV bands, respectively. There is a strong correlation with the 100 μm IRAS flux with an average UV/IR ratio of 300 photons cm$^{-2}$ s$^{-1}$ sr$^{-1}$ Å$^{-1}$ (MJy sr$^{-1}$)$^{-1}$ in the FUV band and 220 photons cm$^{-2}$ s$^{-1}$ sr$^{-1}$ Å$^{-1}$ (MJy sr$^{-1}$)$^{-1}$ in the NUV but with significant variations over the sky. In addition to the large scale distribution of the diffuse light, we note a number of individual features including bright spots around the hot stars Spica and Achernar.

## *Introduction*

The diffuse sky background extends and is well understood over most of the electromagnetic spectrum with contributions from a range of galactic and extragalactic sources (Primack et al. 2008). Observations have, however, been patchy in both distribution and quality in the ultraviolet (UV) and it is only recently that large scale observations of the diffuse UV sky are emerging (reviewed by Murthy 2009; Bowyer 1991; and Henry 1991). The bulk of the radiation is well correlated with the infrared sky background (Sasseen et al. 1995; Schiminovich et al. 2001) consistent with an origin in stellar radiation scattered by interstellar dust. A baseline of about 200 ph cm$^{-2}$ s$^{-1}$ sr$^{-1}$ Å$^{-1}$ (Henry 1991 and references therein) is clearly present at high galactic latitudes, some part of which is due to other galaxies (Armand, Milliard, & Deharveng 1994, Xu et al. 2005), but with other, more speculative, sources proposed as well (Henry, Murthy, & Sujatha 2010). On top of the overall galactic background are a few regions of intense emission where bright stars light up nearby interstellar dust (Murthy & Sahnow 2004).

The first attempts to map the large scale distribution of the diffuse sky background were from the S2/68 telescope aboard the TD-1 satellite (Morgan, Nandy, & Thompson 1976), the ELS photometer on the D2B/Aura satellite (Maucherat-Joubert, Cruvellier & Deharveng 1978) and a photometer on the Dynamics Explorer (DE-1) satellite (Fix, Craven, & Frank 1989; Shalima, Murthy, & Fix 2010). However, none of these instruments were optimized for measurements of the diffuse background and the first instrument specifically designed to measure the diffuse background was the Narrowband Ultraviolet Imaging Experiment for Wide-Field Surveys (NUVIEWS) which observed about 25% of the sky (Schiminovich et al. 2001) followed by SPEAR/FIMS (Edelstein et al. 2006) with coverage of 80% of the sky.

The launch of the Galaxy Evolution Explorer (*GALEX*) in 2003 provided a new platform for diffuse UV observations. In the 7 years since its launch, *GALEX* has obtained over 30,000 observations covering 75% of the sky in two ultraviolet bands (far-ultraviolet, FUV: 1350 – 1750 Å and near-ultraviolet, NUV: 1750 – 2850 Å). Although a few of the observations were as long as several tens of thousands of seconds, allowing a study of small scale variations in the diffuse UV background (Sujatha et al. 2009), most were only a few hundred seconds in length necessitating binning over the entire field. The primary intent of this work is to present the diffuse background as observed from *GALEX*, the most complete and sensitive survey of the diffuse background to date.

## *Observations*

The *GALEX* instrument consists of a Ritchey-Chrétien telescope with a diameter of 50 cm for the primary mirror. Light from the sky is separated by a dichroic onto two detectors (FUV and NUV) with a spatial resolution of 3" – 5" over a 1.25° field of view. The instrument and its primary mission has been described by Martin et al. (2005). The primary data products (Morrissey et al. 2007) from a given observation are images of the given field in each of the two bands and a merged point source catalog. A background file is also created for each band by zeroing out the stars in the full image, binning over 192" squares and then interpolating back to the full *GALEX* resolution (see Morrissey et al. 2007 for a full description).

The median of the individual backgrounds (from the background file) at the position of each of the stars in the *GALEX* merged catalog for a given observation is archived in a publicly accessible database at the Multimission Archive at the Space Telescope Science Institute (MAST) from whence we have retrieved them. We have tested these values against our independently derived sky backgrounds in two sets of deep observations (Sujatha et al. 2009, 2010) and find good agreement. However, we have also found that the background varies over the *GALEX* field at scales of a few arcminutes or less with a standard deviation of 50 – 200 ph cm$^{-2}$ s$^{-1}$ sr$^{-1}$ Å$^{-1}$, depending on the structure in the field. This should be considered as the uncertainty in our values with the possibility of greater deviations in regions with more structure.

The *GALEX* data include not just the astrophysical background but also airglow and zodiacal light. Both depend on the date and time of the observation and will vary over the case of a long observation which may consist of several exposures spread over months or even years. The airglow in either band is between about 200 - 300 ph cm$^{-2}$ s$^{-1}$ sr$^{-1}$ Å$^{-1}$ (Boffi et al. 2007; Sujatha et al. 2009). We have estimated the uncertainty in the airglow to be about 50 ph cm$^{-2}$ s$^{-1}$ sr$^{-1}$ Å$^{-1}$. The airglow is a strong function of the local time of the observation (Sujatha et al. 2009) which was not easily recoverable for our observations. Zodiacal light, sunlight scattered by interplanetary dust, may be estimated using its observed distribution in the optical scaled by the solar spectrum (Leinert et al. 1998). Although often taken to be unity, the color index (relative to the solar spectrum) was found to increase with distance from the ecliptic plane by Murthy et al. (1990) introducing an additional uncertainty in the derived backgrounds. The level of the zodiacal light depends on the angle from the Sun (the elongation angle) and the ecliptic latitude of the observation and ranges from 300 - 1000 ph cm$^{-2}$ s$^{-1}$ sr$^{-1}$ Å$^{-1}$ in the NUV with an estimated uncertainty of about 20%. There is no contribution (< 5 ph cm$^{-2}$ s$^{-1}$ sr$^{-1}$ Å$^{-1}$) of the zodiacal light to the FUV band because the solar flux vanishes below about 2000 Å.

We have tabulated the derived backgrounds with the airglow and zodiacal light subtracted for both the FUV and NUV observations in Table 1 and plotted them in an Aitoff map in Fig. 1. Several regions, most obviously the Galactic plane, but also other bright regions such as the Magellanic Clouds and Orion, were excluded from this survey because of concerns for the safety of the GALEX detectors. In addition, the sky coverage in the FUV and the NUV bands is not identical because of intermittent power problems with the FUV detector. We estimate the total uncertainty in the derived backgrounds to be on the order of 300 ph cm$^{-2}$ s$^{-1}$ sr$^{-1}$ Å$^{-1}$, due to a combination of the uncertainties in the foreground and the spatial variability of the diffuse galactic background on arcminute scales.

## *Results*

Readily apparent in Fig. 1 is the drop-off in both the FUV and the NUV radiation from the Galactic plane, following a cosecant distribution with galactic latitude (Fig. 2). The slope is 545 photons cm$^{-2}$ s$^{-1}$ sr$^{-1}$ Å$^{-1}$ in the FUV band and 433 photons cm$^{-2}$ s$^{-1}$ sr$^{-1}$ Å$^{-1}$ in the NUV bands, close to that found by Wright (1992) in his reanalysis of the Dynamics Explorer (DE-1) data of Fix et al. (1989). The comparable slope for the 100 μm emission from the Infrared Astronomy Satellite (IRAS) is 2.6 MJy sr$^{-1}$ (Boulanger & Pérault 1987), implying an average FUV/IR ratio of 210 photons cm$^{-2}$ s$^{-1}$ sr$^{-1}$ Å$^{-1}$ (MJy sr$^{-1}$)$^{-1}$ and an NUV/IR ratio of 167 photons cm$^{-2}$ s$^{-1}$ sr$^{-1}$ Å$^{-1}$ (MJy sr$^{-1}$)$^{-1}$. Of course, many phenomena originating in the Galaxy would result in a cosecant law but the origin of the majority of the UV emission in dust scattered light is shown by its correlation with the 100 μm IRAS emission (Fig. 3). The FUV/IR ratio has a slope of 302 photons cm$^{-2}$ s$^{-1}$ sr$^{-1}$ Å$^{-1}$ (MJy sr$^{-1}$)$^{-1}$, a y-intercept of 106 photons cm$^{-2}$ s$^{-1}$ sr$^{-1}$ Å$^{-1}$ and a correlation coefficient of 0.82 with the NUV/IR ratio having a slope of 220 photons cm$^{-2}$ s$^{-1}$ sr$^{-1}$ Å$^{-1}$ (MJy sr$^{-1}$)$^{-1}$ with an intercept of 193 photons cm$^{-2}$ s$^{-1}$ sr$^{-1}$ Å$^{-1}$ and a correlation coefficient of 0.72. In both cases we have restricted the fit to those locations where the optical depth is less than 0.7 where the UV and IR radiation are both due to emission from optically thin layers of dust. As the column density of the dust increases, the optical depth increases rapidly in the UV while the medium remains optically thin in the IR. Hence, the UV flux saturates while the 100 μm flux continues to increase linearly. In addition, local effects become important as relatively thin clouds in front of massive molecular clouds dominate the UV while the IR arises primarily in the background cloud as observed eg. in Ophiuchus (Sujatha et al. 2005) and the Coalsack (Sujatha et al. 2007). It should be noted here that the offsets we find are dependent on the zero points of the IRAS data as well as the GALEX data, either of which may be wrongly estimated.

Observed UV/IR ratios in the literature range from cluster around the 200 – 300 ph cm$^{-2}$ s$^{-1}$ sr$^{-1}$ Å$^{-1}$ (MJy sr$^{-1}$)$^{-1}$ mark (Sasseen & Deharveng 1996; Perault et al. 1991; Wright 1992) but with smaller values (< 200 photons cm$^{-2}$ s$^{-1}$ sr$^{-1}$ Å$^{-1}$ (MJy sr$^{-1}$)$^{-1}$) found in FAUST observations of individual regions (Sassen & Deharveng 1996; Haikala et al. 1995) and by Schiminovich et al. (2001) who found ratios of between 50 and 100 photons cm$^{-2}$ s$^{-1}$ sr$^{-1}$ Å$^{-1}$ (MJy sr$^{-1}$)$^{-1}$ in their survey of 25% of the sky. We do find significant variation in our observed FUV/IR and NUV/IR ratios (Fig. 4) with a high of about 1000 photons cm$^{-2}$ s$^{-1}$ sr$^{-1}$ Å$^{-1}$ (MJy sr$^{-1}$)$^{-1}$ near sources of UV radiation (Orion and the individual stars of Spica and Achernar) but detailed modeling of the individual regions is required for an understanding of the process.

The primary astrophysical contributor to both the FUV and NUV bands is scattered light from interstellar dust and the two are well-correlated (Fig. 5) with a correlation coefficient of 0.97. The ratio between the two (FUV/NUV) is 1.41 with an offset of -158 photons cm$^{-2}$ s$^{-1}$ sr$^{-1}$ Å$^{-1}$, where the offset may represent the uncertainty in the foreground emission in each band. Despite this outstanding correlation, there is still considerable structure in the FUV/NUV ratio (Fig. 6). This may reflect other contributions to the FUV band, in particular, such as molecular hydrogen fluorescence (Martin, Hurwitz, & Bowyer 1990; Sujatha et al. 2009) or emission from excited lines such as CIV (Korpela et al. 2006; Park et al. 2009). These contributors may be as high as 20% or more of the total observed signal.

Several individual features are apparent in Fig. 1 and the other maps of the sky, amongst the most prominent of which are the extended dust halos around the bright early-type stars Spica (l, b = 316.2°, 50.8°) and Achernar (l, b = 290.8°, -58.8°). The halo around Spica might be expected as thermal emission from a nearby cloud is seen in the infrared (Zagury, Jones & Boulanger 1998) while an immense faint Hα emission nebula extends as far as 18° from the star (Reynolds 1985). No similar extended nebula has been associated with Achernar but that star is a fast rotator with circumstellar emission extending far from the star and it is possible that we are seeing reflection from this medium. These stars and many of the other features seen in the diffuse sky will well repay further study.

We now have a wealth of UV observations of the dust scattered radiation at both low and high spatial resolution and, for the first time, our data are better than the models. The UV background is largely due to the reflection of starlight from interstellar dust grains; the stellar radiation which is not reflected is absorbed and is then reemitted as thermal emission in the infrared. There has been considerable progress in modeling the dust emission in the infrared (eg. Arendt et al. 1998); the onus is now to create unified models for dust scattering and the subsequent thermal emission which include multiple scattering and clumping. Such models have been applied to FAUST data by Witt et al. (1997) who found an albedo of $0.45 \pm 0.05$ and a phase function asymmetry factor (g = $<\cos\theta>$) of $0.68 \pm 0.10$ and and NUVIEWS (Schiminovich et al. 2001) data who found an albedo of $0.45 \pm 0.05$ and g = $0.77 \pm 0.1$. Although we have derived similar values when we applied such models to our own data, we have found too many deviations in individual targets to place much faith in the results.

## *Conclusions*

Our main focus in this work has been in presenting the data. We found that existing models would fit the gross dependence on Galactic latitude found here but could not match with individual targets, probably because of the complexity of the interstellar medium (Witt & Gordon 2000).

Our future work will focus on two areas. The first is in developing (or adapting) more realistic models of the dust distribution including both multiple scattering and clumpiness in the gas and dust and applying them to the all-sky GALEX data. The UV background is sensitive to the geometry of the sources (stars) and the scatterers (dust) and may depend on local effects, unlike the infrared which is more dependent on the total column density. The second is in a more careful study of smaller areas of the sky such as around Spica where the dust halo is seen and the poles where we can further constrain the extragalactic flux.

The GALEX data have proven to be one of the most sensitive probes of the diffuse background and have yielded a wealth of data that will prove a challenge to analyze and interpret. Of particular interest will be its future exploitation with better models and in conjunction with spectroscopic data such as provided by SPEAR/FIMS. We look forward to a new era in studies of the UV diffuse radiation.

Acknowledgements: We thank Patrick Morrissey for help in understanding the processing of the GALEX data. An anonymous referee has greatly helped in extending and clarifying this work. The research leading to these results received funding from the Indian Space Research Organization through the Space Science Office, the Department of Science and Technology under the Young Scientists Program and the National Aeronautics and Space Administration through the Maryland Space Grant program. This research is based on observations made with the NASA's GALEX program, obtained from the data archive at depending on the detailed geometry of the distribution of the source stars and the scattering/absorbing interstellar dust. Note that the optical depth in Figs. 3a and 3b refer to the optical depth at 1530 and 2310 Å, respectively.

Fig. 1. The diffuse sky at 1530 Å (1a) and at 2310 Å (1b). The Galactic centre is at the plot centre and axis lines are at 30° intervals with the North Galactic Pole at the top. A color scale in units of photons cm$^{-2}$ s$^{-1}$ sr$^{-1}$ Å$^{-1}$ is on the right hand side of the maps. Black regions are those with no data, primarily because of the fear of damaging the detectors.

Fig. 2. The latitude dependence of the FUV (2a) and NUV (2b) radiation. Cosecant laws of slope 545 photons cm$^{-2}$ s$^{-1}$ sr$^{-1}$ Å$^{-1}$ (FUV: solid line) and 433 photons cm$^{-2}$ s$^{-1}$ sr$^{-1}$ Å$^{-1}$ (NUV: dashed line) are over-plotted. Fig 2c and 2d show the same data but this time with the FUV (Fig. 2c) and NUV (Fig. 2d) fluxes plotted as a function of the cosecant of the Galactic latitude. In both of the latter plots, points below the Galactic equator are plotted in red while points above the equator are plotted in blue. This distribution is consistent with the idea that the bulk of the UV radiation is starlight scattered from interstellar dust grains.

Fig. 3. The FUV/IR (3a) and NUV/IR (3b) correlations. The slope (solid line) of the FUV/IR correlation is 302 photons cm$^{-2}$ s$^{-1}$ sr$^{-1}$ Å$^{-1}$ (MJy sr$^{-1}$)$^{-1}$ with a y-intercept of 106 photons cm$^{-2}$ s$^{-1}$ sr$^{-1}$ Å$^{-1}$ and a correlation coefficient is 0.82 ($\tau < 0.7$). The slope (dashed line) of the NUV/IR correlation is 220 photons cm$^{-2}$ s$^{-1}$ sr$^{-1}$ Å$^{-1}$ (MJy sr$^{-1}$)$^{-1}$ with a y-intercept of 193 photons cm$^{-2}$ s$^{-1}$ sr$^{-1}$ Å$^{-1}$ and a correlation coefficient is 0.72 ($\tau < 0.7$). The reason for the strong correlation is that the UV radiation that is not scattered is absorbed and heats the grains, the energy being re-emitted as IR radiation. The large scatter at large optical depth (top scale) is because at substantial optical depths the radiative transfer becomes complex,

Fig. 4. The FUV/IR (4a) and NUV/IR ratio (4b) in units of photons cm$^{-2}$ s$^{-1}$ sr$^{-1}$ Å$^{-1}$ (MJy sr$^{-1}$)$^{-1}$.

Fig. 5. The correlation between FUV (1530 Å) and NUV (2310 Å). The solid line represents the FUV/NUV ratio with a slope of 1.41 and an offset of -158 photons cm$^{-2}$ s$^{-1}$ sr$^{-1}$ Å$^{-1}$. The very tight correlation suggests that the primary contributor to both bands is dust scattered starlight.

Fig. 6. The FUV/NUV ratio. Variations in the ratio may indicate regions of line and molecular emission.

TABLE 1

Diffuse Flux Observed by GALEX

| Gal. Longitude | Gal. Latitude | Total FUV | Total NUV | FUV | NUV |
|---|---|---|---|---|---|
| | | (ph cm$^{-2}$ s$^{-1}$ sr$^{-1}$ Å$^{-1}$) | | (ph cm$^{-2}$ s$^{-1}$ sr$^{-1}$ Å$^{-1}$) | |
| 261.46 | -40.15 | 582.28 | 1117.73 | 393.93 | 559.12 |
| 240.38 | -49.19 | 457.28 | 1001.7 | 335 | 427.72 |
| 239.27 | -47.61 | 454.9 | 979 | 331.01 | 404.52 |
| 239.09 | -48.54 | 445.51 | 918.88 | 226.52 | 284.49 |
| 238.91 | -49.47 | 513.77 | 1052.24 | 395.05 | 474.03 |
| 234.85 | -48.33 | 581.6 | 1099.56 | 372.83 | 460.44 |
| 234.61 | -49.23 | 550.13 | 1026.22 | 338.87 | 385.22 |
| 85.16 | 70.66 | 425.86 | 1029.44 | 341.85 | 429.14 |
| 240.54 | -48.25 | 446.88 | 968.14 | 322.04 | 394.74 |
| 237.82 | -47.88 | 442.03 | 915.66 | 228.14 | 280.89 |
| 160.67 | 51.23 | 361.17 | 1021.7 | 271.87 | 291.5 |
| 82.3 | 75.56 | 420.98 | 1133.76 | 317.09 | 455.49 |
| 152.81 | 52.36 | 394.26 | 1094.62 | 273.36 | 375.22 |
| 76.17 | 69.4 | 384.56 | 902.32 | 170.87 | 239.07 |
| 236.34 | -48.12 | 473.02 | 944.91 | 261.68 | 307.96 |
| 87.88 | 72.24 | -1 | 1016.11 | -1 | 427.24 |
| 236.12 | -49.02 | 518.1 | 997.82 | 304.24 | 358.97 |

Notes: Table 1 is published in its entirety in the electronic edition of the Astrophysical Journal Suppl. Series. A portion is shown here for guidance regarding its form and content. A "-1" in the cell implies that no valid data is available or that cell. The coordinates are those of the center of the 1.25° field; the total FUV and total NUV are the median value of the background (including airglow and zodiacal light) over the field; FUV and NUV have had the airglow and zodiacal light subtracted.

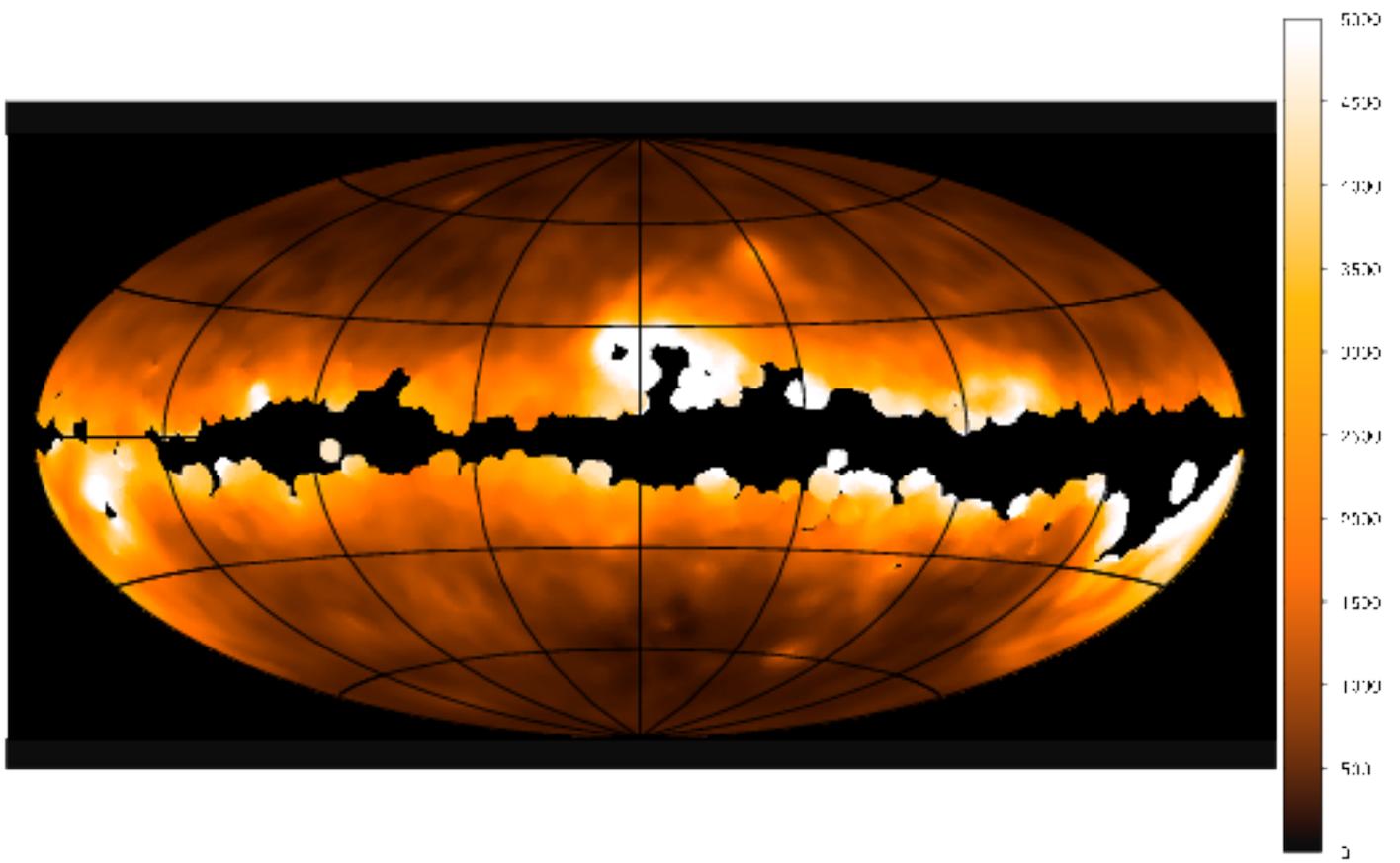

Figure 1a: FUV

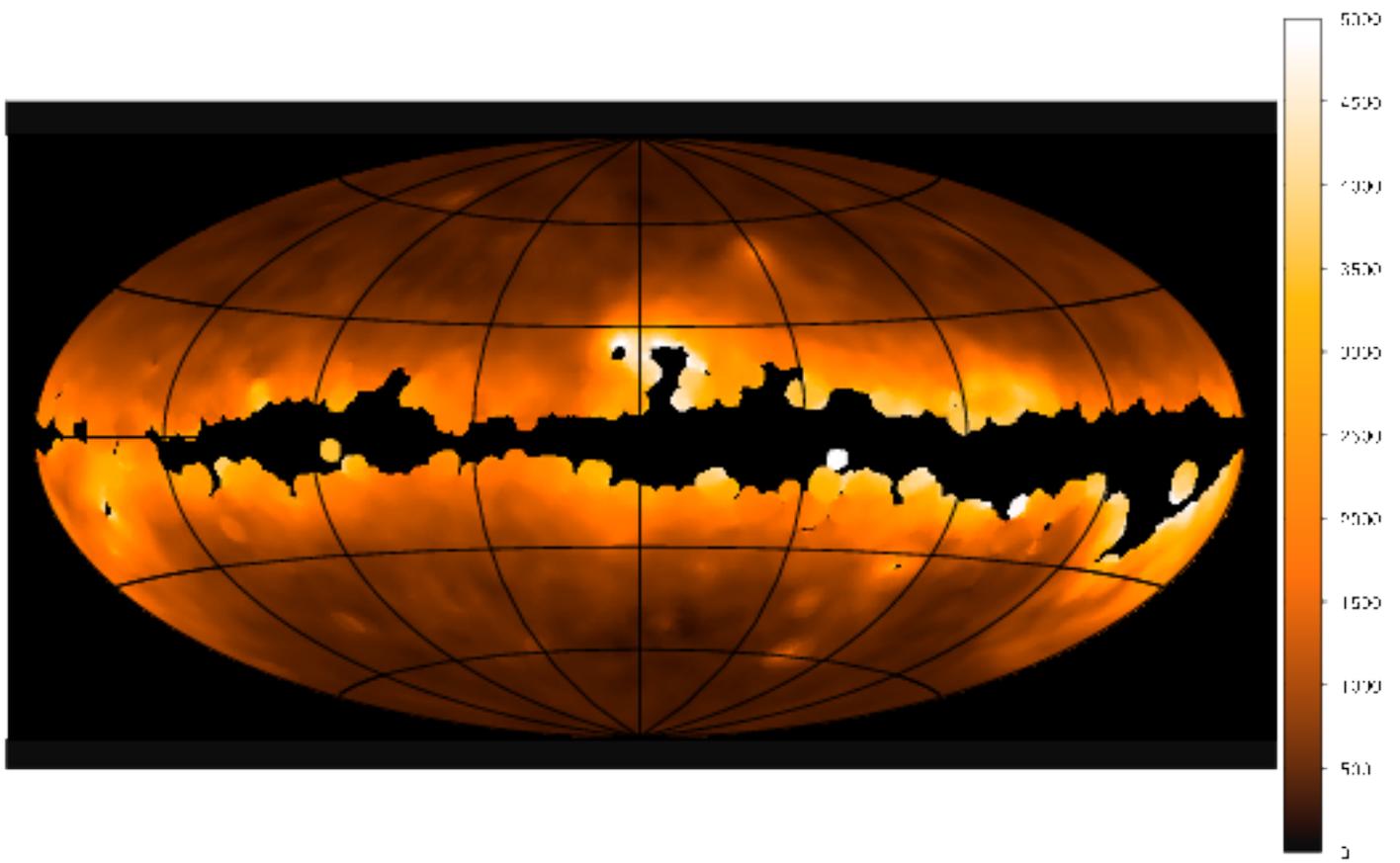

Figure 1b: NUV

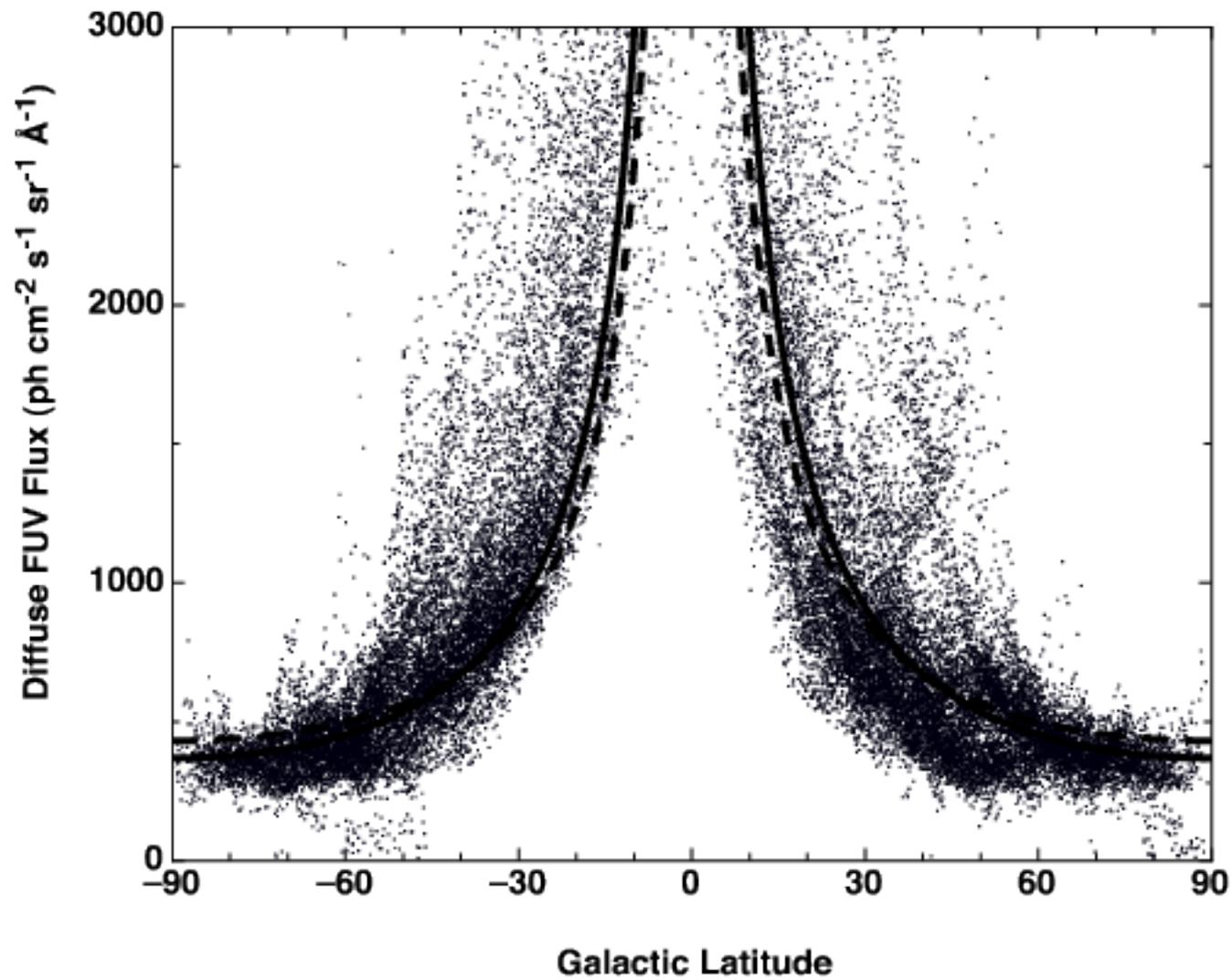

Figure 2a

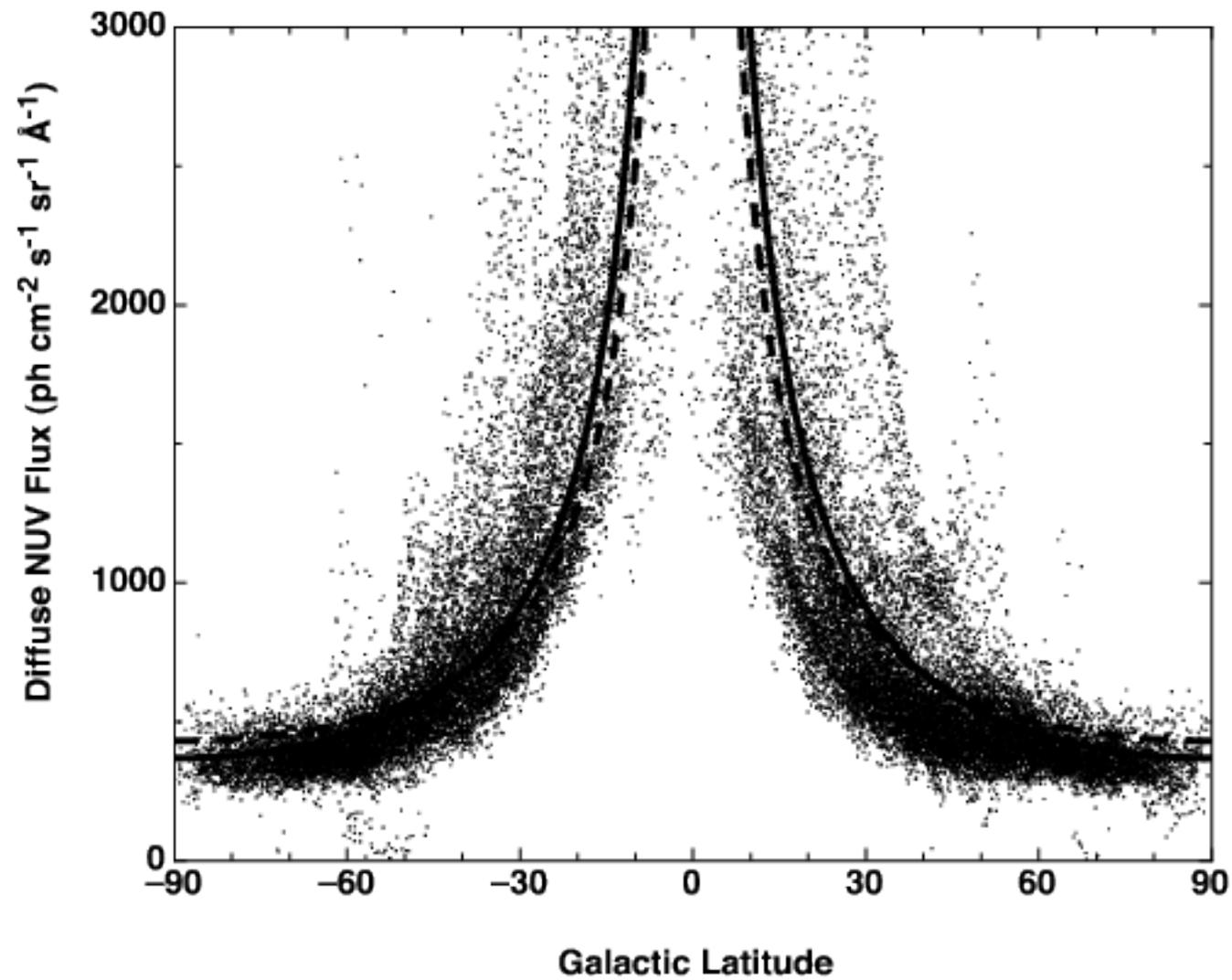

Figure 2b

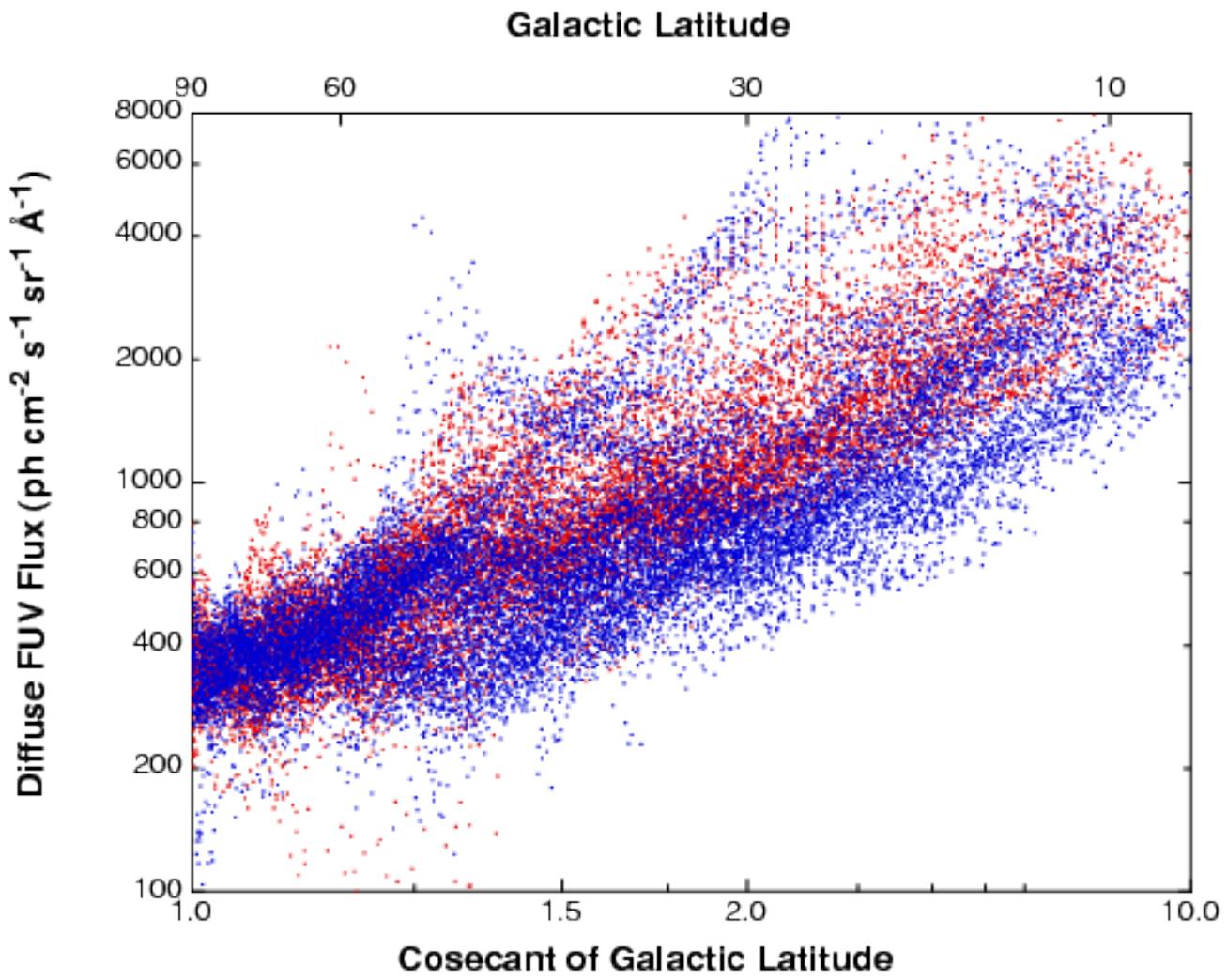

Fig. 2c

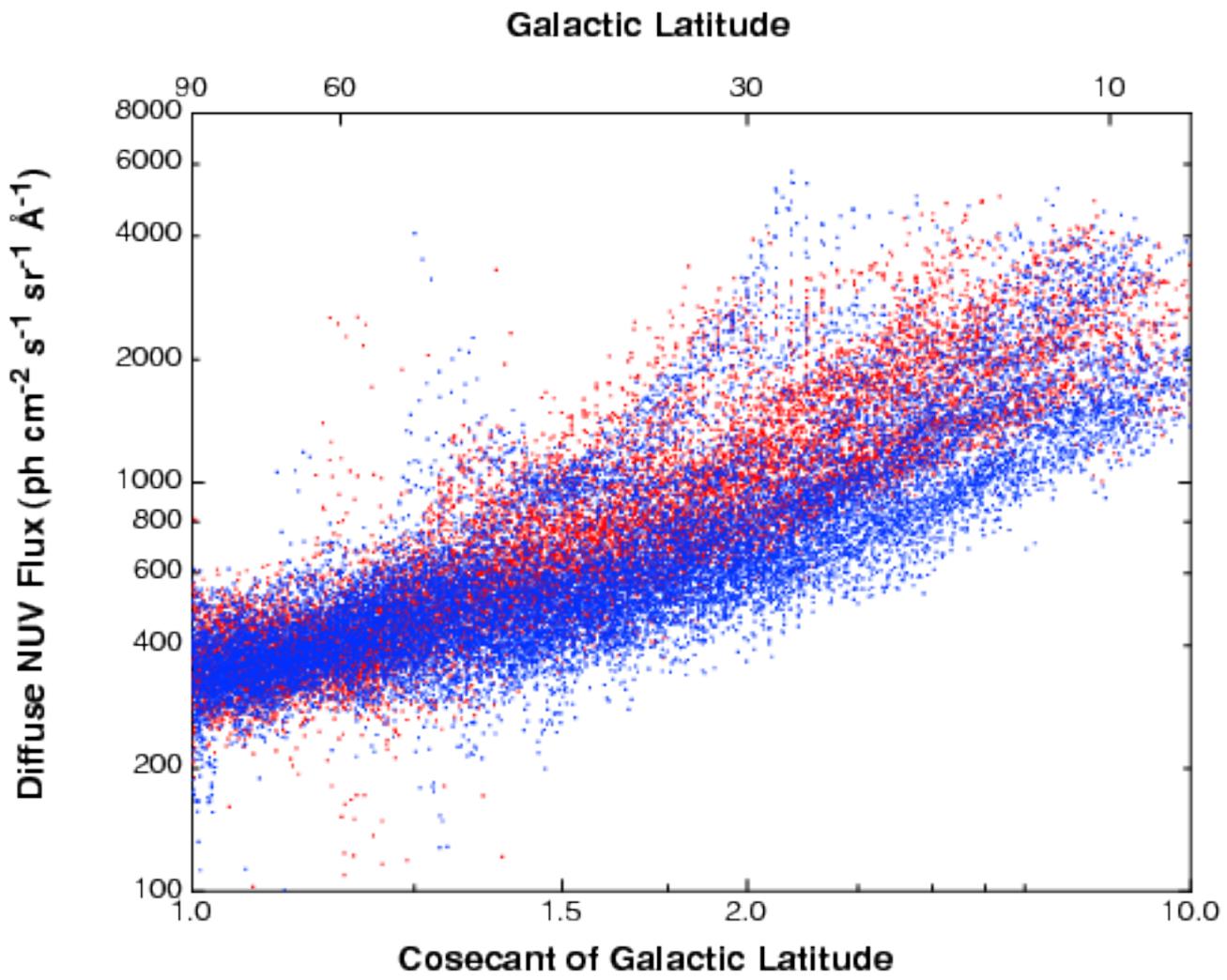

Fig. 2d

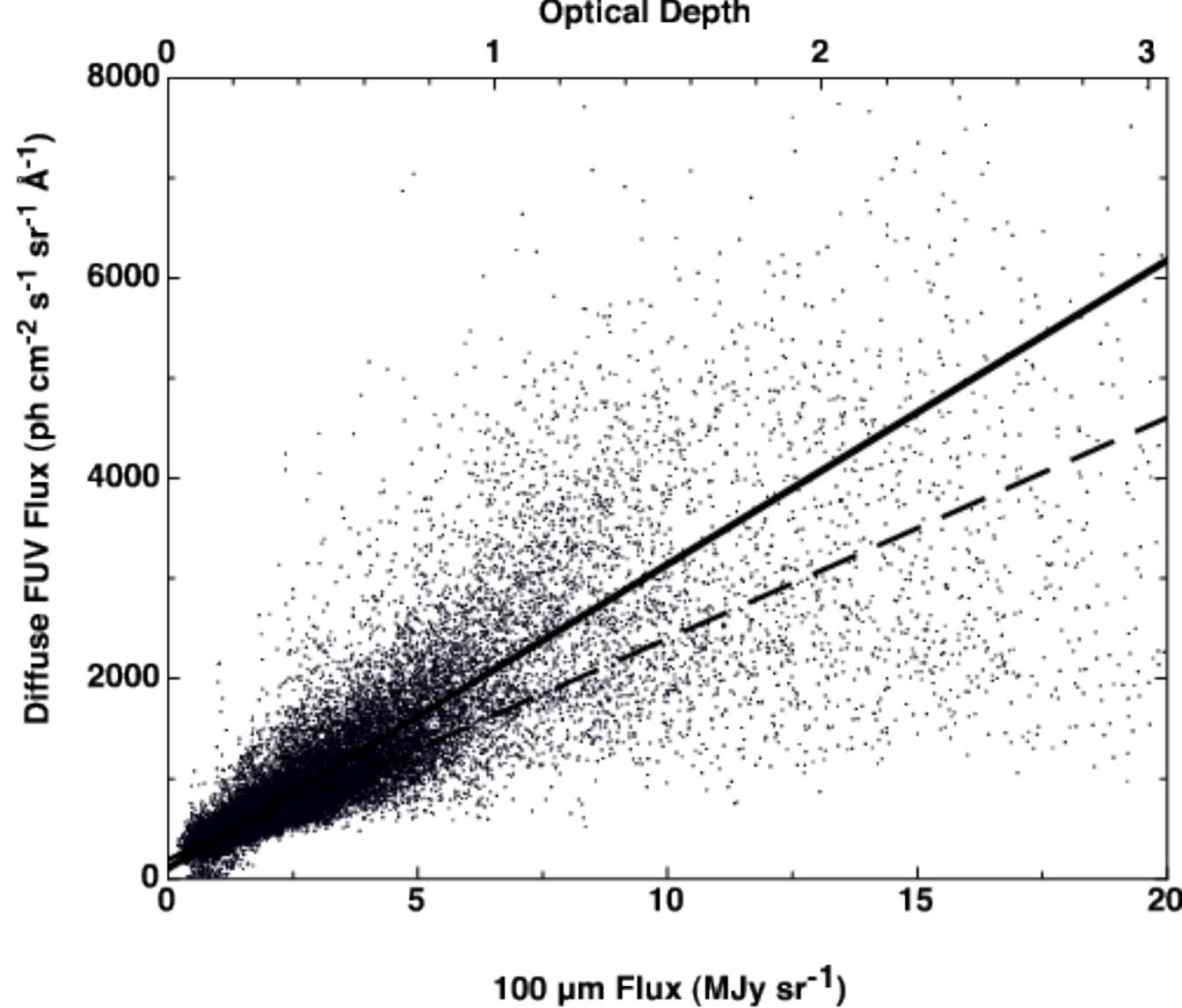

Figure 3a

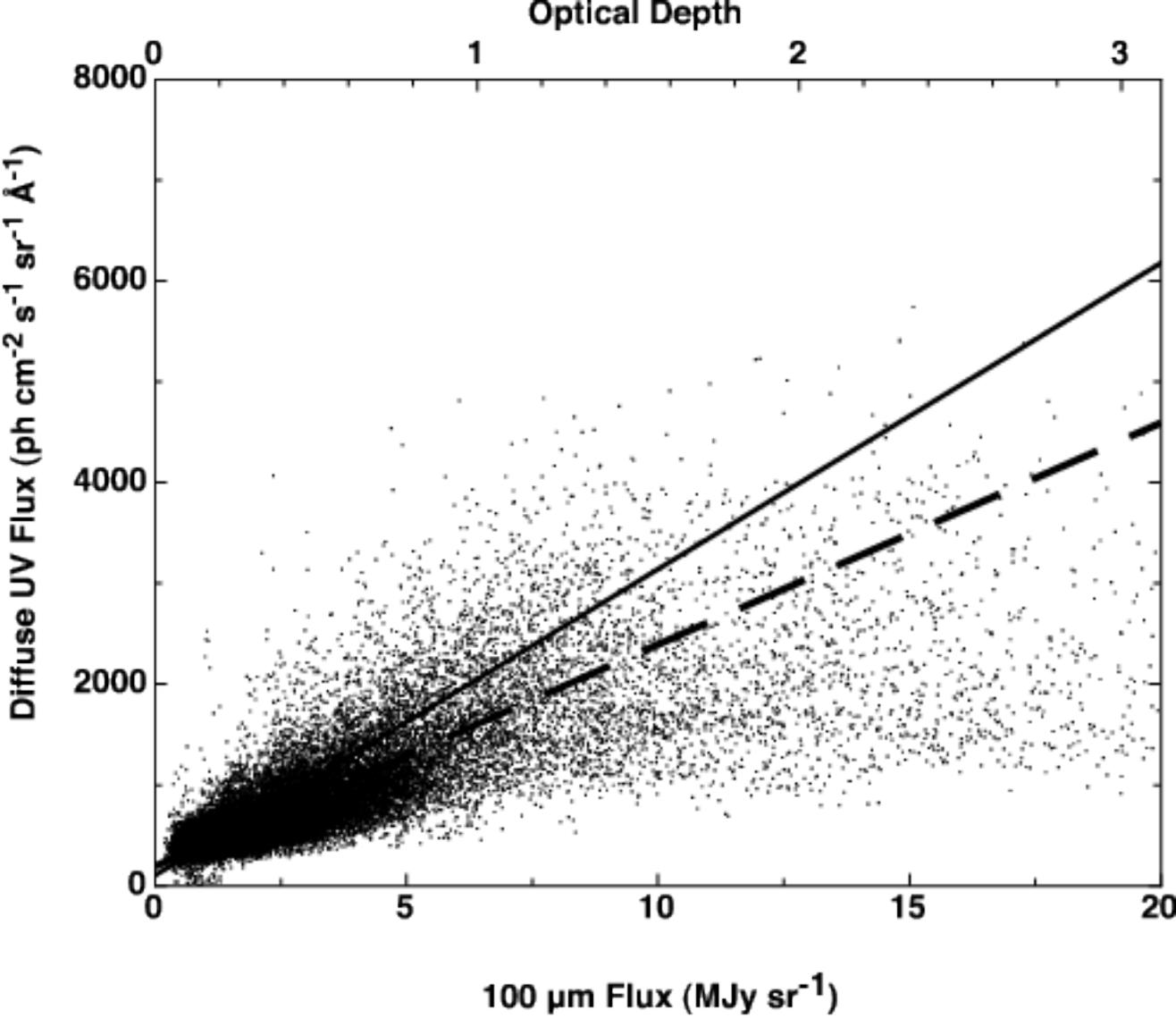

Figure 3b

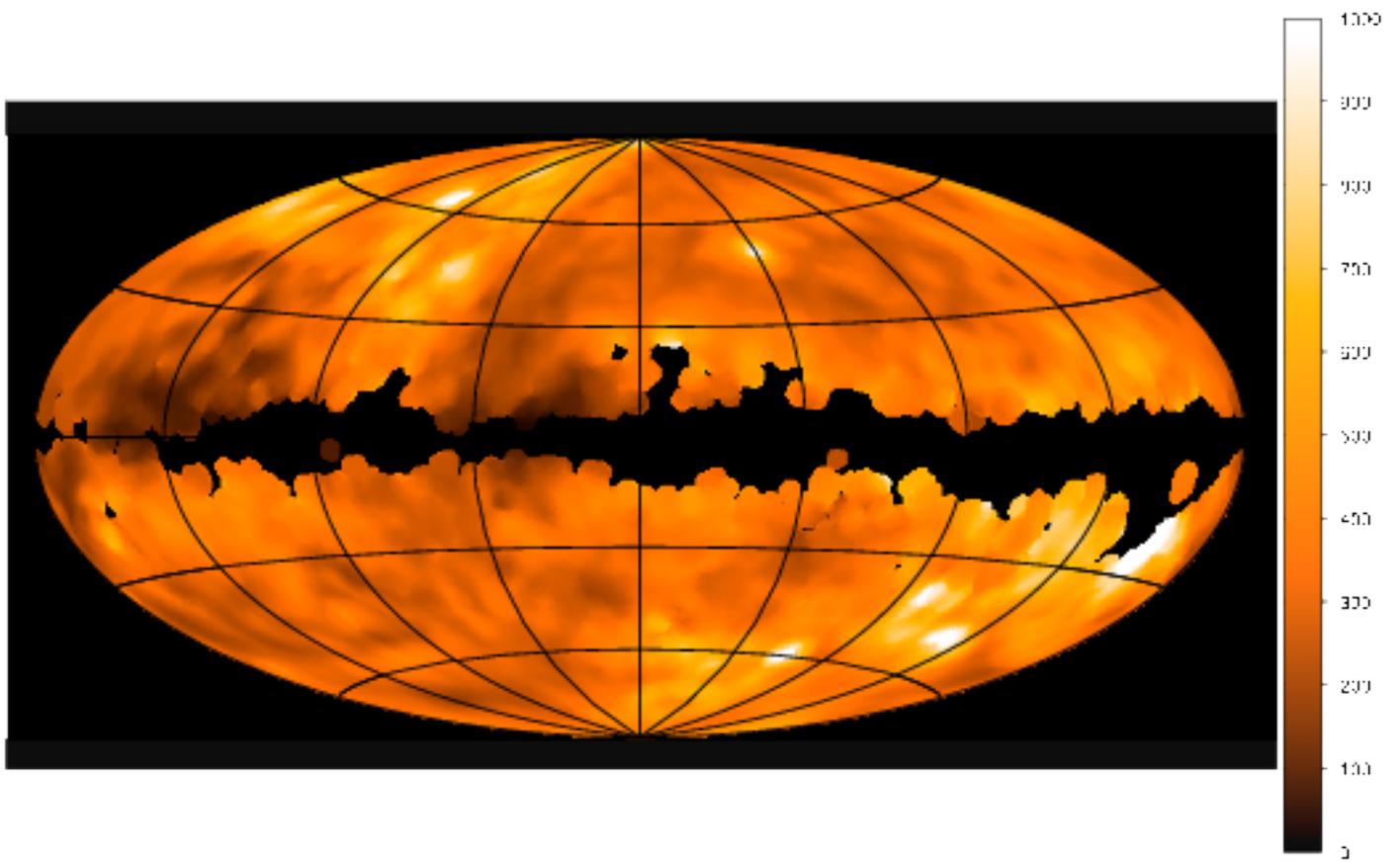

Figure 4a: FUV/IR

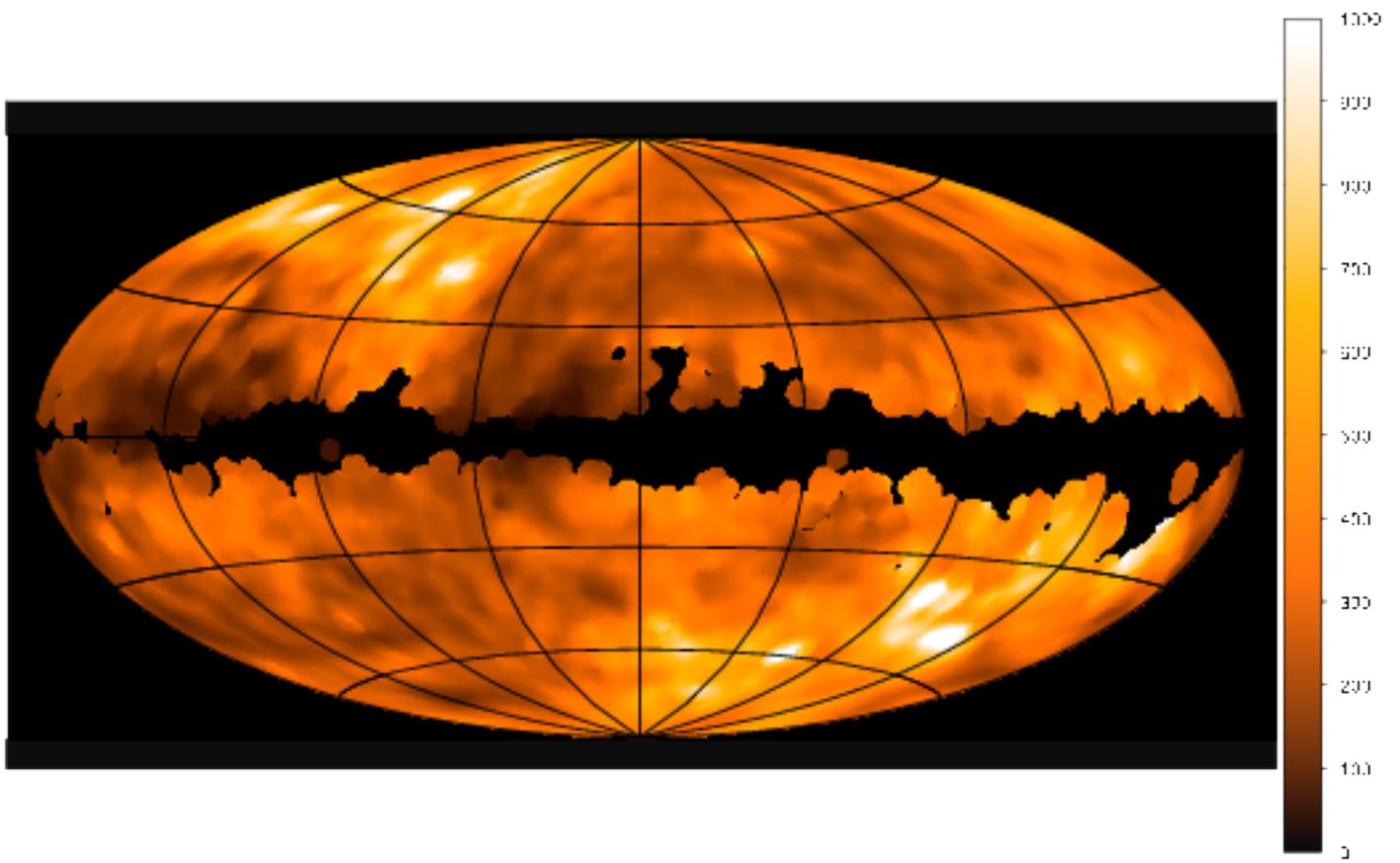

Figure 4h: NUVIB

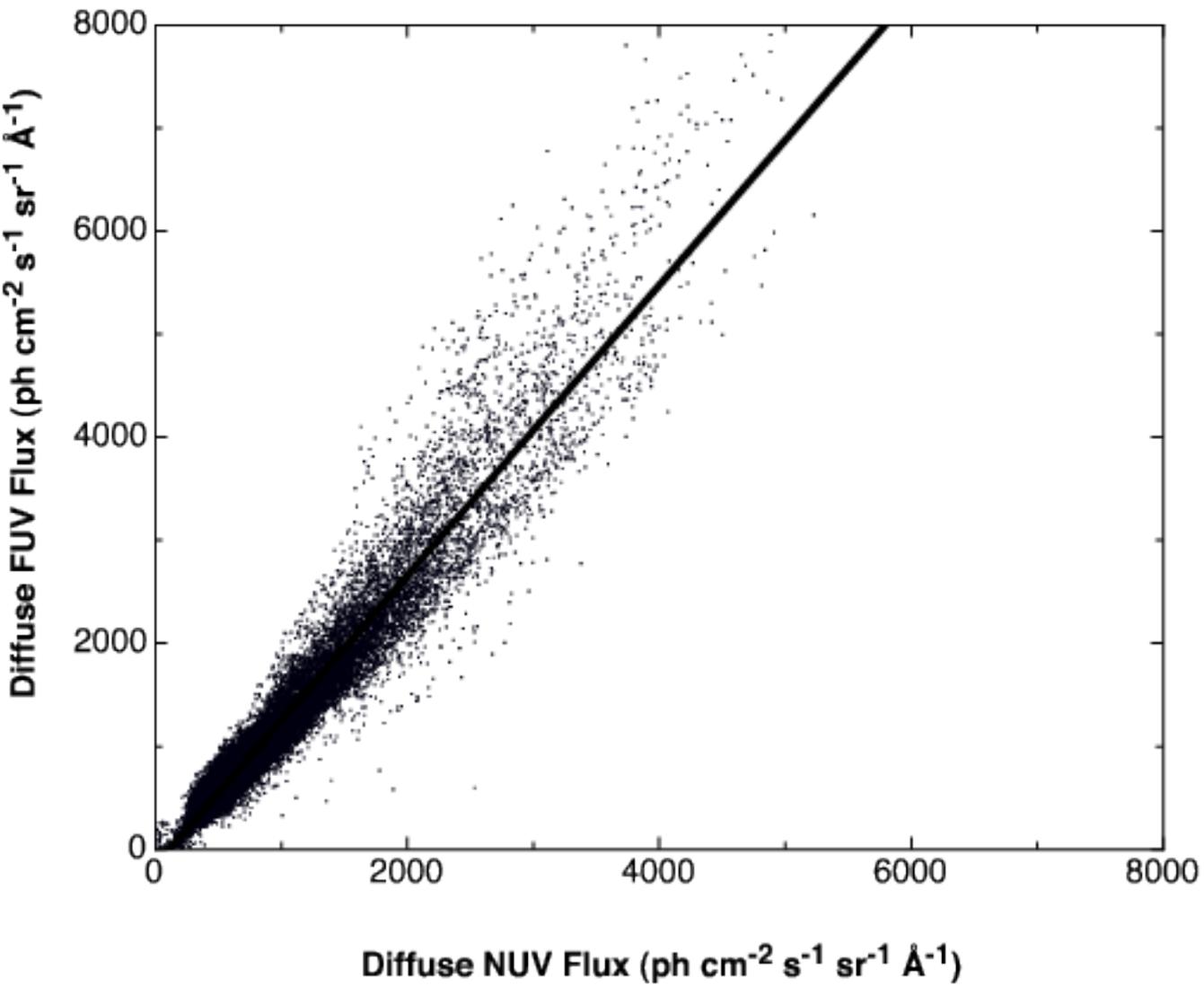

Fig. 5

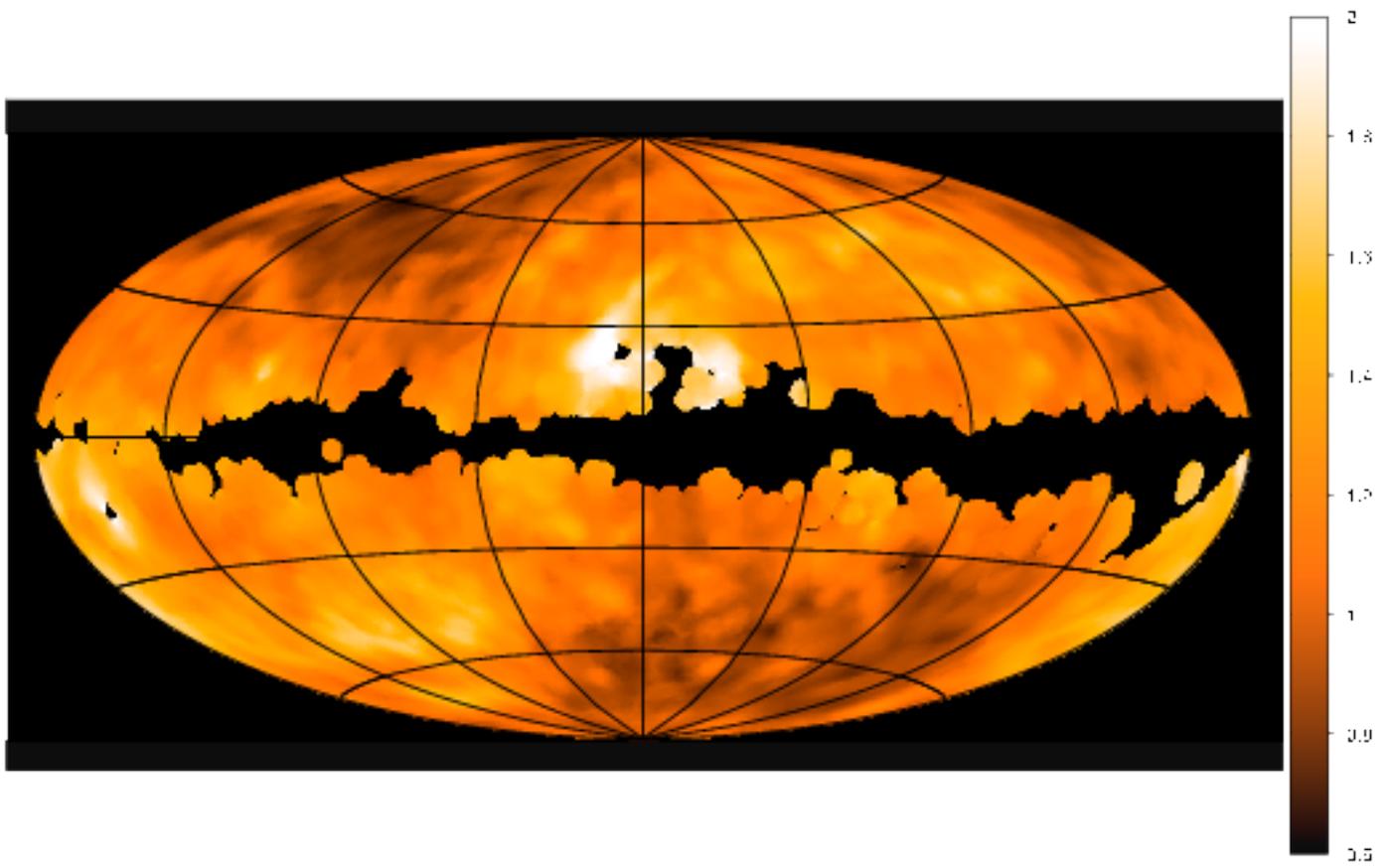

Fig. 6

Figure 6: FUV/NUV